\documentclass[graybox]{svmult}

\usepackage[numbers]{natbib}
\usepackage{mathptmx}      
\usepackage{helvet}         
\usepackage{courier}       
\usepackage{type1cm}       
\usepackage{makeidx}         
\usepackage{graphicx}        
\usepackage[bottom]{footmisc}
\usepackage{graphicx}
\usepackage{subcaption}
\usepackage{eucal}
\usepackage{textcomp}
\usepackage{xcolor}
\usepackage{booktabs} 
\usepackage[ruled, linesnumbered]{algorithm2e}
\usepackage{balance}
\usepackage{tikz}
\usepackage{amsmath}
\usetikzlibrary{trees}
\usepackage{lipsum,adjustbox}
\usetikzlibrary{positioning}
\usetikzlibrary{mindmap,trees}
\usepackage{verbatim}
\usepackage{multirow}
\usepackage[T1]{fontenc}
\usepackage{makecell}
\usepackage{amssymb}
\usepackage{hyperref}
\usepackage{amsmath}
\usepackage{relsize}
\usepackage{stmaryrd}
\usepackage{breqn}
\usepackage[shortlabels, inline]{enumitem}
\usepackage{url}

\usepackage{color, colortbl}
\definecolor{Gray}{gray}{0.9}
\definecolor{airforceblue}{rgb}{0.36, 0.54, 0.66}
\definecolor{aliceblue}{rgb}{0.94, 0.97, 1.0}
\definecolor{alizarin}{rgb}{0.82, 0.1, 0.26}
\definecolor{amber}{rgb}{1.0, 0.75, 0.0}
\definecolor{amber(sae/ece)}{rgb}{1.0, 0.49, 0.0}
\definecolor{bronze}{rgb}{0.8, 0.5, 0.2}
\definecolor{battleshipgrey}{rgb}{0.52, 0.52, 0.51}
\definecolor{bole}{rgb}{0.47, 0.27, 0.23}
\definecolor{bulgarianrose}{rgb}{0.28, 0.02, 0.03}
\definecolor{cadet}{rgb}{0.33, 0.41, 0.47}
\definecolor{ceil}{rgb}{0.57, 0.63, 0.81}
\definecolor{cerulean}{rgb}{0.0, 0.48, 0.65}
\definecolor{charcoal}{rgb}{0.21, 0.27, 0.31}
\definecolor{coolblack}{rgb}{0.0, 0.18, 0.39}
\definecolor{coolgrey}{rgb}{0.55, 0.57, 0.67}
\definecolor{darkcandyapplered}{rgb}{0.64, 0.0, 0.0}
\definecolor{darkbrown}{rgb}{0.4, 0.26, 0.13}
\definecolor{darkcerulean}{rgb}{0.03, 0.27, 0.49}
\definecolor{darkgray}{rgb}{0.66, 0.66, 0.66}
\definecolor{darkjunglegreen}{rgb}{0.1, 0.14, 0.13}
\definecolor{darktaupe}{rgb}{0.28, 0.24, 0.2}
\definecolor{davy\'sgrey}{rgb}{0.33, 0.33, 0.33}
\definecolor{frenchblue}{rgb}{0.0, 0.45, 0.73}
\definecolor{almond}{rgb}{0.94, 0.87, 0.8}
\definecolor{beaublue}{rgb}{0.74, 0.83, 0.9}
\definecolor{beige}{rgb}{0.96, 0.96, 0.86}
\definecolor{bisque}{rgb}{1.0, 0.89, 0.77}
\definecolor{black}{rgb}{0.0, 0.0, 0.0}
\definecolor{fluorescentorange}{rgb}{1.0, 0.75, 0.0}
\definecolor{ghostwhite}{rgb}{0.97, 0.97, 1.0}
\definecolor{antiquewhite}{rgb}{0.98, 0.92, 0.84}
\definecolor{LightCyan}{rgb}{0.88,1,1}

\begin{document}

\title*{\centering Trustworthiness of $\mathbb{X}$ Users: A One-Class Classification Approach}
\titlerunning{ Trustworthiness of $\mathbb{X}$ Users}


\author{\centering Tanveer Khan\thanks{The first two authors contributed equally to this work.}, Fahad Sohrab, Antonis Michalas, Moncef Gabbouj}
\institute{Tampere University, Finland. \email{(tanveer.khan, fahad.sohrab, antonios.michalas, moncef.gabbouj)@tuni.fi}}

\authorrunning{Khan \textit{et al.}}
%
%
\maketitle
\abstract{$\mathbb{X}$ (formerly Twitter) is a prominent online social media platform that plays an important role in sharing information making the content generated on this platform a valuable source of information. Ensuring trust on $\mathbb{X}$ is essential to determine the user credibility and prevents issues across various domains. While assigning credibility to $\mathbb{X}$ users and classifying them as trusted or untrusted is commonly carried out using traditional
machine learning models, there is limited exploration about the use of One-Class Classification (OCC) models for this purpose. In this study, we use various OCC models for $\mathbb{X}$ user classification. Additionally, we propose using a subspace-learning-based approach that simultaneously optimizes both the subspace and data description for OCC. We also introduce a novel regularization term for Subspace Support Vector Data Description (SSVDD), expressing data concentration in a lower-dimensional subspace that captures diverse graph structures. Experimental results show superior performance of the introduced regularization term for SSVDD compared to baseline models and state-of-the-art techniques for $\mathbb{X}$ user classification.}

\section{Introduction}
\label{sec:introduction}
Online Social Networks (OSNs) have become an essential tool for modern communication, enabling people to interact, while 
spending significant time on these platforms. It is now becoming an integral part of our lives, and people are using it for different purposes, including connection with friends and family, participation in online communities, brand promotion, 
finding and sharing information, and much more~\cite{gazi2017research}. The most popular OSNs include Facebook, $\mathbb{X}$, Instagram, and LinkedIn. 
This work focuses on $\mathbb{X}$ -- an OSN platform that allows users to share and discover short messages or tweets limited to 280 characters. $\mathbb{X}$ has~310 million active users publishing~500 million tweets per day~\cite{das2019much}. Also, $\mathbb{X}$ has become a valuable tool enabling users to share information with a wide audience quickly and easily. 
It allows users to see tweets relevant to their interests, retweet or like other users' tweets, or post their own tweets. While this makes it easy for $\mathbb{X}$ users to share updates and information with their followers in real-time, it also makes it easier for fake account users to carry out malicious activities such as sharing unverified information~\cite{vosoughi2018spread}. Additionally, it has been observed that fake news spreads more rapidly on $\mathbb{X}$ than real news, 
damaging the reputation and reliability of the $\mathbb{X}$. 
Various techniques have been proposed~\cite{khan2021fake} to tackle the spread of false information, with one approach being the classification of $\mathbb{X}$ users as trusted or untrusted~\cite{zhang2018social}. This classification is of significant importance 
in maintaining the reputation and reliability of the $\mathbb{X}$ platform. For example, identifying a trusted and reliable $\mathbb{X}$ user ensures the continued success and usefulness of the $\mathbb{X}$ platform as a trusted and valuable social media tool.

There are various ways for $\mathbb{X}$ user classification such as using 
Machine Learning (ML)~\cite{ pritzkau2021finding}, and Natural Language Processing (NLP)~\cite{devarajan2023ai}. Among these, ML models have been widely used in various research to classify $\mathbb{X}$ users into different categories based on their profiles, activity, and content of the tweets. The process involves collecting and preprocessing large amounts of data, including user profiles and tweets, and then training an ML model to classify users into different categories based on the available features. The model can then be used to classify new, unseen users. The ML algorithms used for $\mathbb{X}$ user classification are supervised~\cite{asfand2023classifying}, unsupervised~\cite{ahmad2020information}, and semi-supervised~\cite{khan2020trust, khan2021seeing}. Classifying $\mathbb{X}$ users as trusted or untrusted using only ML models can be challenging due to high-dimensional and variable characteristics of big data~\cite{wang2020survey}. Despite the \textit{curse of dimensionality} and the imbalanced nature of the data, the appropriate techniques and models hold the potential to address these challenges successfully. In our approach, we rely on OCC models, where the decision function is inferred using training data from a single class only. It is used when a large amount of data is available for the class of interest but little or no data is available for other classes~\cite{alam2020one}. OCC differs from traditional binary classification models, which are trained using data from both categories. We use a manually labeled dataset obtained from Khan \textit{et al.}~\cite{khan2020trust} research, which involved gathering data for~50,000 $\mathbb{X}$ users, with manual labeling for~1,000 of them. By applying different OCC models to the labeled dataset, our goal is to answer the following \textbf{research questions (RQs)}:

\textbf{\textit{RQ 1}}: How effective is the OCC in accurately identifying political $\mathbb{X}$ users as trusted or untrusted, and what are the comparative strengths and weaknesses among different OCC models in this context?

\textbf{\textit{RQ 2}}: What are the key challenges OCC faces when classifying political users on $\mathbb{X}$, and can the performance of OCC be optimized for political user identification through subspace learning for OCC?

\textbf{\textit{RQ 3}}: Can we encode the relationships between the training data points in a lower-dimensional subspace optimized for OCC while capturing and preserving the local structure of target class data?

\noindent \textit{\textbf{Contributions:}}
The main contributions of this work can be summarized as follows:

\begin{enumerate}[\bfseries C1., leftmargin=0.7cm]
      \item We propose using subspace-learning-based OCC for $\mathbb{X}$ user identification. 
      \item We propose a novel regularizer for Subspace Support Vector Data Description (SSVDD) expressing the concentration of the data in a lower-dimensional subspace that captures different graph structures.
    \item  In the proposed regularization term, any suitable graph can be used to encode the corresponding graph structure, and we evaluate its effectiveness by comparing it with different OCC models.
\end{enumerate}
\subsection{Organization}
The rest of the paper is organized as follows. In~\autoref{sec:preliminaries}, we provide necessary background information about different OCC models. In~\autoref{sec:related_Work}, we provide important published works in the area of $\mathbb{X}$ user credibility, accompanied by a detailed discussion of our proposed approach in~\autoref{sec:methodology}. The data collection and experimental results are presented in~\autoref{sec:exp and eva}. Finally, we conclude the paper in~\autoref{sec:conclusion}.
\section{Preliminaries}
\label{sec:preliminaries}
\label{subsec:onecc}
In ML, OCC refers to an approach to building a model by considering data from a single class only. OCC is appropriate for scenarios 
where it is critical to identify one of the
categories, but the examples from that specific category are scarce or statistically so diverse that they cannot be used during the training process. OCC has found application in different areas, such as early detection of myocardial infection \cite{degerli2022early}, 
rare insect classification \cite{sohrab2020boosting}, and 
credit card fraud detection \cite{zaffar2023credit}. These applications 
present data scarcity challenges from one of the categories to be modeled. 

Among the widely-used OCC approaches, One-class Support Vector Machine (OCSVM) and Support Vector Data Description (SVDD) have been 
proven as powerful data description methods over time. These methods identify the so-called \textit{support vectors} as crucial for determining the decision boundary. In OCSVM, a hyperplane is created to separate the target class in a way that maximizes the distance of the hyperplane from the origin~\cite{scholkopfu2000sv}. The classification of a new data point is determined by its location relative to the hyperplane: if it falls on the positive side, it is considered normal; otherwise, it is flagged as abnormal. SVDD, on the other hand, creates a hyperspherical boundary around the target class data within the original feature space by minimizing the volume of the hypersphere. 

Let us denote the target class training samples to be encapsulated inside a hypersphere by a matrix $\mathbf{X}=[\mathbf{x}_{1},\mathbf{x}_{2},\dots, \mathbf{x}_{N}],\mathbf{x}_{i} \in \mathbb{R}^{D}$, where $N$ is the number of samples and $D$ is dimensionality of data. The formulation of SVDD is expressed as follows:

{\fontsize{9}{8}\selectfont
\begin{align}\label{erfuncSVDD2}
\centering
\min \quad  F(R,\mathbf{a}) = R^2 + C\sum_{i=1}^{N} \xi_i \ \ \ 
\textrm{s.t.} \quad  \|\mathbf{x}_i - \mathbf{a}\|_2^2 \le R^2 + \xi _i, \quad \xi_i \ge 0, \:\: \forall i\in\{1,\dots,N\},
\end{align}}
where $R$ represents the radius, $\mathbf{a}\in \mathbb{R}^{D}$ is the center of the hypersphere, and slack variables $\xi_i,i=1,\dots, N$ are introduced to 
enable the possibility of target data being outliers. The hyperparameter $C>0$ controls the trade-off between the volume of the hypersphere and the presence of data points outside the hypersphere. A test sample is assigned to the positive class if its distance from the center of the hypersphere is equal to or less than the radius $R$.

A distinct category in OCC, Graph Embedded One-Class Classifiers, refers to methods that integrate generic graph structures expressing relevant geometric relationships in their optimization processes.
Graph Embedded One-Class Support Vector Machine (GEOCSVM) is an example that incorporates graph-based information and enhances the traditional OCSVM approach. By leveraging graph information, GEOCSVM compares favorably to the standard OCSVM. In GEOCSVM, the relationship between training patterns can be described locally and globally using a single graph or a combination of fully connected and kNN graphs~\cite{mygdalis2016graph}. Similarly, Graph Embedded Support Vector Data Description (GESVDD) is a type of OCC that combines the SVDD approach with graph-based information. In GESVDD, the graph-based information is incorporated into the optimization process of the SVDD. Like SVDD, GESVDD also creates a hypersphere around the target class data to separate the target class data from the outliers in an OCC problem. However, graph-based information in GESVDD provides additional information that can help to improve the separation of target class data from outliers~\cite{mygdalis2016graph}. Other extensions of graph-based OCC include Graph Embedded Subspace Support Vector Data Description (GESSVDD) \cite{sohrab2023graph} that poses the subspace learning for OCC as a graph embedding problem.

Traditional boundary-based OCC methods primarily find a data description in the given feature space. However, a contemporary paradigm shift is evident in the form of subspace learning-based techniques that not only form a data description but also optimize a subspace simultaneously. A leading technique in this paradigm is the SSVDD \cite{sohrab2018subspace}, which defines a data description along with data mapping to low-dimensional feature space optimized for OCC.
To define a concise representation of the target class, the method repeatedly optimizes 
data mapping and 
data description. 
The optimization function of SSVDD is as follows:
{\fontsize{9}{8}\selectfont
\begin{align}\label{ssvdd}
\centering
\min \quad F(R,\mathbf{a}) = R^2 + C\sum_{i=1}^{N} \xi_i \ \ \
\textrm{s.t.} \quad  \|\mathbf{Qx}_i - \mathbf{a}\|_2^2 \le R^2 + \xi _i,\quad
\xi_i \ge 0, \:\: \forall i\in\{1,\dots,N\},
\end{align}  }
where $\mathbf{Q} \in \mathbb{R}^{d \times D}$ is the projection matrix for mapping the data from the original \textit{D}-dimensional feature space to an optimized lower \textit{d}-dimensional space. In SSVDD, an augmented version of the Lagrangian with a regularization term $\psi$ is optimized: 
\begin{equation}\label{Lang}
L= \sum_{i=1}^{N} \alpha_i  \mathbf{x}_i^\intercal \mathbf{Q}^\intercal \mathbf{Q} \mathbf{x}_i - \sum_{i=1}^{N}\sum_{j=1}^{N} \alpha_i \mathbf{x}_i^\intercal \mathbf{Q}^\intercal \mathbf{Q} \mathbf{x}_j \alpha_j + \beta\psi,
\end{equation}
where $\alpha$ represents the Lagrange multipliers, and $\beta$ is used to control the importance of the regularization term. The regularization term $\psi$ expresses the class variance in the $d$-dimensional space and it is denoted as \begin{equation}
\label{generalconstraintpsi} 
\psi = \text{Tr}(\mathbf{Q}\mathbf{X}\boldsymbol{\lambda}\boldsymbol{\lambda}^\intercal 
 \mathbf{X}^\intercal\mathbf{Q}^\intercal),
\end{equation} 
where Tr($\cdot$) is the trace operator and $\lambda \in \mathbb{R}^N$ is a vector used to select the contribution of certain data points in the optimization process, leading to different variants of SSVDD. The different variants are as follows.

\begin{itemize}
    \item SSVDD$\psi1$: In this variant, the regularization term becomes obsolete and is not used during the data description.
    \item SSVDD$\psi2$: In this case, all the training samples describe the class variance in the regularization term.
    \item SSVDD$\psi3$: In this case, the samples belonging to the boundary and outside the boundary are used in the regularization term.
    \item SSVDD$\psi4$: In this variant, only the support vectors that belong to the class boundary are used to describe the class variance in the regularization term. 
\end{itemize}

The selection of different data instances in the regularization term is carried out by replacing the $\lambda$ value accordingly with the $\alpha$ values. The updating of the projection matrix $\mathbf{Q}$ is carried out by utilizing the gradient of \eqref{Lang}, expressed as:
\begin{align}\label{Quopdate}
\mathbf{Q}\leftarrow \mathbf{Q}-\eta\Delta L.
\end{align}
Here, $\eta$ denotes the learning rate parameter. This work primarily focuses on subspace learning-based OCC and proposes a graph-based regularization for SSVDD. 
\section{Related Work} 
\label{sec:related_Work}
A lot of research has looked into different aspects of $\mathbb{X}$, such as bot detection, analysis of the spread of fake news, and assessing the credibility of $\mathbb{X}$ users. 
Bots can be helpful for tasks such as posting information about news and providing assistance during emergencies, etc~\cite{haustein2016tweets}, but some bots can be used for malicious purposes such as influencing public opinion or spreading malware~\cite{fu2018combating}. Hence, identifying bots is vital for $\mathbb{X}$ to enforce its platform terms and conditions. Hence, researchers have proposed different methods~\cite{rodriguez2020one} to create accurate models for bot detection.

Apart from bot detection, another important area of research is the detection of fake news, which is rampant -- tends to be retweeted faster than true ones~\cite{meyers2020fake}. 
Various ML models, particularly of the supervised classification, have been used for fake news detection~\cite{faustini2020fake}. For example, Hassan \textit{et al.}~\cite{hassan2017toward} extracted features from the sentences and used a support vector machine to detect fake news. 
Despite the popularity of the topic 
there has been limited progress in fake news detection. This is partly due to the ongoing controversy surrounding the term `fake news' and the lack of a universally accepted definition thereof~\cite{meyers2020fake}. Nevertheless, several works have delved into fake news detection by assessing the credibility of tweets or classifying the $\mathbb{X}$ users as trusted or untrusted. 
Another study proposes an automated ranking technique to evaluate tweet credibility. Gupta \textit{et al.} worked on assigning a credibility score to each tweet. Another interesting work in this domain is the work conducted by Tanveer \textit{et al.}~\cite{khan2020trust, khan2021seeing}, presented a model that analyzes $\mathbb{X}$ users, assigning a score to each user based on their social profile, tweet credibility, and h-index score.
While there has been considerable research in this domain, it is important to note that only a limited number of studies utilize an OCC to classify $\mathbb{X}$ user as trusted or untrusted. This underscores a critical gap in the existing body of knowledge. Adopting OCC becomes particularly valuable when the task involves identifying a specific category with limited or diverse training instances. 
\section{Methodology}
\label{sec:methodology}
\begin{figure}[h]
\centering
\includegraphics[width=11.5cm]{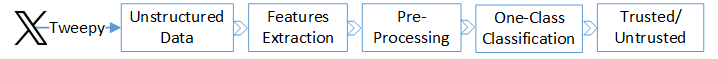}
\caption{One-class classification 
categorizes $\mathbb{X}$ platform users as trusted or untrusted}
\label{fig:pipeline}
\end{figure}

This research aims to develop a regularization strategy for training OCC models, specifically tailored for identifying political $\mathbb{X}$ users, categorizing them as either trusted or untrusted, as shown in~\autoref{fig:pipeline}. For this reason, we use the manually labeled dataset of~1000 political $\mathbb{X}$ users from the paper~\cite{khan2020trust}.
For each user, a unique profile is created, containing various features. 
Some are basic features extracted for each $\mathbb{X}$ user linked to their account. More specifically, these features are 
\begin{enumerate*}[label=(\roman*)]
    \item Number of friends, \item Number of followers, \item Number of retweets, \item Number of likes, \item URLs, \item Lists, \item Status and \item Mention by others.
\end{enumerate*}

The basic features are used to calculate more advanced features like a social reputation score, an h-index score, a sentiment score, and tweet credibility. Below, we provide a brief description of these advanced features:

\begin{itemize}
    \item Social reputation score: It 
    provides the number of users interested in the updates of an $\mathbb{X}$ user. 
    \item H-index score: The h-index is used to measure how impactful an $\mathbb{X}$ user is. 
    This is measured by considering the number of likes and retweets of a $\mathbb{X}$ user.
    \item Sentiment score: The tweets of a $\mathbb{X}$ user are classified as positive, negative, and neutral, 
    based on which 
    sentiment score is assigned to each $\mathbb{X}$ user.
    \item Tweet 
    credibility: It is calculated by considering the retweet ratio, liked ratio, URL ratio, user hashtag ratio, and original content ratio. 
    \item Influence score: The influence score of a $\mathbb{X}$ user is calculated by considering the social reputation, h-index score, sentiment score, and tweet credibility.
\end{itemize}
Details on calculating influence scores from basic features and using advanced features are beyond this paper's scope. For more information, refer to the previous article on this topic \cite{khan2020trust}. All 
political $\mathbb{X}$ users are classified as trusted or untrusted based on social reputation, tweet credibility, sentiment score, h-index score, and influence score. All 
$\mathbb{X}$ accounts with abusive and harassing tweets, a low social reputation, h-index, and influence score are grouped 
as untrusted users, while those who are more reputable among users with a high h-index score, 
more credible tweets and a 
high influence score are grouped 
as trusted users. 

Having a dataset for political $\mathbb{X}$ users as either trusted or untrusted based on various criteria, we then focus on inferring a model based on using information only from trusted users. We train different OCC models, including SVDD, ESVDD, OCSVM, SSVDDr$\psi_{1}$, SSVDDr$\psi_{2}$, SSVDDr$\psi_{3}$, SSVDDr$\psi_{4}$ GEOCSVM and GESVDD. We 
also propose a novel regularization term for SSVDD. The newly proposed regularization term considers the graph information, which measures the concentration of the data in a lower-dimensional subspace and captures the essential features of the training set while 
preserving the local structure of the data. The proposed regularization term is defined as
\begin{equation}
\label{Graphconstraintpsi} 
\gamma = \text{Tr}(\mathbf{Q}\mathbf{X}\mathbf{L}_x\mathbf{X}^\intercal\mathbf{Q}^\intercal),
\end{equation} 
where $\mathbf{L}_x$ is the Laplacian matrix of the graph. The subscript $x$ denotes the adopted graph type. The Laplacian is defined as
\begin{align}\label{graph} 
\mathbf{L}_x=\mathbf{D}_x-\mathbf{A}_x,\;\;\;
[\mathbf{D}_x]_{ii}=\sum_{j\neq i}[\mathbf{A}_x]_{ij},\forall i\in \{1,\dots,N\},
\end{align}
where $\mathbf{D}_x$ is the degree matrix and $\mathbf{A}_x\in \mathbb{R}^{N\times N}$ serves as the graph's weight matrix. In what follows, we drop the subscript X for notation simplicity.

We investigated the three different graph Laplacians in the proposed regularization term $\gamma$. In the first experiment, we exploit the local geometric information by employing k-Nearest Neighbor (kNN) and setting the Laplacian matrix to
\begin{equation}\label{L_kNN}
\mathbf{L}_{kNN} =\mathbf{D}_{kNN}-\mathbf{A}_{kNN},
\end{equation}
where $[\mathbf{A}_{kNN}]_{ij}=1,$ if $\mathbf{x}_{i}\in \mathcal{N}_j$ or $\mathbf{x}_{j}\in \mathcal{N}_i$ and 0, otherwise. 
$\mathcal{N}_i$ denotes the nearest neighbors of $\mathbf{x}_i$. Adjusting the $k$ numbers of neighbors in kNN allows the neighborhoods $\mathcal{N}_i$ to be defined accordingly. In the second experiment, we use within-cluster Laplacian information.

\begin{equation}
\mathbf{L}_{w}=\mathbf{I}-\sum_{c=1}^{\mathcal{C}}\frac{1}{N_c}\mathbf{1}_c\mathbf{1}_c^{T}, \label{eq:sw}
\end{equation}
where $\mathbf{I}$ is an identity matrix, $\mathcal{C}$ denotes the total numbers of clusters, $\mathbf{1}$ is a vector of ones, $N_c$ is the total number of instances belonging to cluster $c$ and
$\mathbf{1}_c$ represents a vector with ones corresponding to instances that belong to cluster $c$ and zeros elsewhere. In the third experiment, we use the between-cluster scatter information:
\begin{equation}
\mathbf{L}_{b}=\sum_{c=1}^{\mathcal{C}}N_c \left(\frac{1}{N_{c}}\mathbf{1}_{c}-\frac{1}{N}\mathbf{1}\right)\left(\frac{1}{N_{c}}\mathbf{1}_{c}-\frac{1}{N}\mathbf{1}^{T}\right).\label{eq:sb}
\end{equation}

In this paper, we denote the three variants of the proposed regularization strategies for SSVDD as SSVDD$\gamma L_{kNN}$, SSVDD$\gamma L_w$, and SSVDD$\gamma L_b$, respectively. For non-linear data description, we employed non-linear projection trick (NPT) \cite{kwak2013nonlinear}. NPT is equivalent to employing the widely recognized kernel trick while enabling the use of the method's linear variant.
The kernel matrix is obtained as
\begin{equation}\label{RBFkernel}
\mathbf{K}_{ij} = \exp  \left( \frac{ -\| \mathbf{x}_{i} - \mathbf{x}_{j}\|_2^2 }{ 2\sigma^2 } \right),
\end{equation} 
where $\sigma$ is a hyperparameter scaling the distance between $\mathbf{x}_i$ and $\mathbf{x}_j$. We followed similar steps for non-linear data description as adapted in recent variants and extensions of SSVDD \cite{sohrab2020ellipsoidal, sohrab2023newton}.

\section{Experimental Results and Model Evaluation}
\label{sec:exp and eva}
To extract the features from $\mathbb{X}$ and generate the dataset, we used Python~3.5. The Python script was executed locally on a machine with the following configuration: Intel Core~i7,~2.80*8~GHZ,~32GB, Ubuntu~16.04 LTS~64 bit. For the training and evaluation of the OCC models, we switched to Matlab and
performed the experiments on Intel(R) Xeon(R) CPU~E5-2650 v3~2.30GHz~64GB RAM.
We provide the open-source implementation of our work on Github\footnote{\url{https://github.com/fahadsohrab/xssvdd}}.

A comprehensive set of evaluating metrics is reported over the test set to compare different OCC models. Accuracy (Accu) provides the ratio of correctly classified instances to the total number of instances, True Positive Rate (TPR) represents the proportion of positive instances correctly classified, while True Negative Rate (TNR) indicates the ratio of true negatives to the total number of negative samples. Precision (Pre) measures the proportion of instances classified as positive that are truly positive, and the F1-score is defined as the harmonic mean of precision and TPR. Additionally, Geometric Mean 
is employed to discern the best-performing parameters on training set, calculated as square root of product of TPR and TNR.
\subsection{Preprocessing the data}
In this work, we chose to analyze the $\mathbb{X}$ account of~1,000 politicians~\footnote{\url{https://zenodo.org/records/7014109}} and the main reason for evaluating the profiles of politicians is their intrinsic potential to influence public opinion as their content originates and exists in a sphere of political life, which is, unfortunately, often surrounded by controversial events and outcomes. 
We selected~70 percent of the data for training and~30 percent for testing. The train and test sets are randomly selected by keeping the proportions of the two classes similar to the collected dataset. We 
perform the random selection five times; hence, we use five different train-test sets for the experiments to check the robustness of the OCC methods. We normalize the data by subtracting
the mean and dividing it by the Standard Deviation (STD). These are 
both computed using only the target class samples from the training set. During the training, a~5-fold cross-validation technique is used over the training set to select the hyperparameters of the models. More details on the hyperparameters can be found in the \href{https://github.com/fahadsohrab/xssvdd}{GitHub}.
\subsection{Results and discussions}
In~\autoref{table: occperformance}, we report the average performance measures of various OCC methods on the five data splits of the dataset. The classifiers are divided into two categories: linear OCC and non-linear OCC. In ~\autoref{table: occperformance}, we also report the STD of evaluating metrics for the linear and non-linear methods over the five data splits of the dataset. 

Considering the GM values, the linear OCC generally have lower performance measures than non-linear OCC. This indicates that non-linear OCC are more adept at correctly predicting both positive and negative classes than linear OCC. For example, in non-linear OCC, SSVDD$\gamma L_{kNN}$ achieves the highest GM value, which is~0.80, surpassing the~0.64 obtained by SSVDD$\gamma L_{w}$, a linear OCC. Conversely, non-linear SSVDD$\psi1$ OCC achieve the lowest GM value which is 0.43, 
as opposed to~0.19 recorded by ESVDD linear OCC.

Regarding Accu, most non-linear OCC models consistently outperform their linear counterparts. As shown in~\autoref{table: occperformance}, the highest Accu, reaching~0.80, is achieved by non-linear OCC SSVDD$\gamma L_{kNN}$. In contrast, three linear OCC models -- SSVDD$\gamma L_{w}$, SSVDD$\gamma L_{b}$ and SSVDD$\gamma L_{kNN}$ -- received a slightly lower Accu of~0.74. The lowest Accu among non-linear OCC models is 0.48, attributed to OCSVM, while the linear OCSVM achieves 0.44. For the other evaluation metrics, Pre and F1 remain stable, while TPR consistently remains high, indicating the model's effectiveness in identifying positive instances.

To summarize, linear OCSVM has the lowest Accu (0.44) and F1-score (0.41) among all classifiers, while SVDD and ESVDD have very low TNR and GM values in linear cases. Linear SSVDD classifiers with regularization terms $\psi1$ and $\psi4$ have similar performance measures and are somewhat better than OCSVM, SVDD, and ESVDD. GEOCSVM has the highest Accu (0.78) and GM (0.78) scores, indicating its superior performance in identifying positive and negative instances.

The superior performance of non-linear OCC models in terms of GM and Accu can be attributed to the inherent complexity of the data distribution. Non-linear classifiers are more flexible in capturing intricate relationships and patterns within the data, especially when the decision boundary is non-linear.

Examining the SSVDD variants and their performance metric, GM, concerning the regularization term $\psi$, reveals that $\psi3$ yields the most favorable outcomes for both linear and non-linear classifiers followed closely by $\psi2$, then $\psi1$, with $\psi4$ performing the least effectively. The superiority of SSVDD$\psi3$ can be attributed to its consideration of samples inside and outside the class boundary during the training in the regularization term, providing a more comprehensive understanding of the class variance. Conversely, SSVDD$\psi4$ performs poorly as it only considers support vectors on the class boundary in the regularization term, potentially missing crucial information about class distribution.

\begin{table}[h]
\centering
\footnotesize
\caption{Measuring the performance of linear and non-linear OCC models averaged over five test splits with ± STD} 
\label{table: occperformance}
\setlength{\tabcolsep}{3pt}
\begin{tabular}{|l|llllll|}
\hline
 & Accu & TPR & TNR & Pre & F1 & GM \\
\hline
\multirow{2}{*}{} & \multicolumn{6}{c|}{Linear OCC} \\
\cline{2-7}
SSVDD$\psi1$ & 0.68 ± 0.02 & \textbf{0.98} ± 0.01 & 0.27 ± 0.05 & 0.65 ± 0.01 & 0.78 ± 0.01 & 0.51 ± 0.05 \\
SSVDD$\psi2$ & 0.69 ± 0.03 & 0.97 ± 0.02 & 0.31 ± 0.07 & 0.66 ± 0.02 & 0.78 ± 0.02 & 0.54 ± 0.06 \\
SSVDD$\psi3$ & 0.73 ± 0.06 & 0.97 ± 0.02 & 0.41 ± 0.16 & 0.70 ± 0.06 & 0.81 ± 0.03 & 0.62 ± 0.12 \\
SSVDD$\psi4$ & 0.69 ± 0.03 & \textbf{0.98} ± 0.01 & 0.29 ± 0.08 & 0.66 ± 0.02 & 0.79 ± 0.02 & 0.53 ± 0.07 \\
OCSVM & 0.44 ± 0.10 & 0.34 ± 0.10 & 0.59 ± 0.36 & 0.59 ± 0.13 & 0.41 ± 0.04 & 0.39 ± 0.16 \\
SVDD & 0.58 ± 0.01 & 0.96 ± 0.01 & 0.05 ± 0.03 & 0.59 ± 0.01 & 0.73 ± 0.00 & 0.22 ± 0.06 \\
ESVDD & 0.57 ± 0.01 & 0.96 ± 0.02 & 0.04 ± 0.02 & 0.58 ± 0.00 & 0.72 ± 0.01 & 0.19 ± 0.06 \\
SSVDD$\gamma L_{kNN}$ & 0.74 ± 0.04 & 0.97 ± 0.01 & 0.41 ± 0.12 & 0.70 ± 0.04 & 0.81 ± 0.02 & 0.63 ± 0.09 \\
SSVDD$\gamma L_{b}$ & 0.74 ± 0.07 & \textbf{0.98} ± 0.01 & 0.42 ± 0.19 & 0.71 ± 0.07 & \textbf{0.82} ± 0.04 & 0.63 ± 0.14 \\
SSVDD$\gamma L_{w}$ & 0.74 ± 0.02 & 0.97 ± 0.01 & 0.42 ± 0.04 & 0.70 ± 0.02 & 0.81 ± 0.01 & 0.64 ± 0.03 \\
\hline
\multirow{2}{*}{} & \multicolumn{6}{c|}{Non-Linear OCC} \\
\hline
SSVDD$\psi1$ & 0.66 ± 0.09 & 0.88 ± 0.15 & 0.35 ± 0.39 & 0.68 ± 0.12 & 0.75 ± 0.04 & 0.43 ± 0.31 \\
SSVDD$\psi2$ & 0.70 ± 0.08 & 0.80 ± 0.16 & 0.55 ± 0.35 & 0.75 ± 0.14 & 0.75 ± 0.04 & 0.61 ± 0.19 \\
SSVDD$\psi3$ & 0.74 ± 0.09 & 0.80 ± 0.11 & 0.67 ± 0.31 & 0.80 ± 0.12 & 0.79 ± 0.06 & 0.70 ± 0.20 \\
SSVDD$\psi4$ & 0.65 ± 0.09 & 0.85 ± 0.19 & 0.36 ± 0.41 & 0.69 ± 0.13 & 0.73 ± 0.06 & 0.40 ± 0.33 \\
OCSVM & 0.48 ± 0.04 & 0.53 ± 0.04 & 0.40 ± 0.12 & 0.56 ± 0.04 & 0.54 ± 0.02 & 0.45 ± 0.06 \\
SVDD & 0.71 ± 0.12 & 0.84 ± 0.11 & 0.53 ± 0.43 & 0.76 ± 0.16 & 0.78 ± 0.05 & 0.56 ± 0.34 \\
ESVDD & 0.56 ± 0.02 & 0.58 ± 0.17 & 0.54 ± 0.26 & 0.66 ± 0.09 & 0.60 ± 0.07 & 0.52 ± 0.04 \\
GEOCSVM & 0.78 ± 0.02 & 0.74 ± 0.06 & 0.83 ± 0.06 & 0.86 ± 0.03 & 0.79 ± 0.03 & 0.78 ± 0.01 \\
GESVDD & 0.70 ± 0.12 & 0.61 ± 0.27 & 0.84 ± 0.13 & 0.87 ± 0.09 & 0.67 ± 0.24 & 0.68 ± 0.17 \\
SSVDD$\gamma L_{kNN}$ & \textbf{0.80} ± 0.05 & 0.76 ± 0.11 & \textbf{0.85} ± 0.08 & \textbf{0.88} ± 0.05 & 0.81 ± 0.06 & \textbf{0.80} ± 0.05 \\
SSVDD$\gamma L_{b}$ & 0.73 ± 0.09 & 0.77 ± 0.10 & 0.68 ± 0.21 & 0.78 ± 0.10 & 0.77 ± 0.07 & 0.71 ± 0.12 \\
SSVDD$\gamma L_{w}$ & 0.78 ± 0.02 & 0.84 ± 0.07 & 0.70 ± 0.08 & 0.80 ± 0.03 & \textbf{0.82} ± 0.03 & 0.76 ± 0.02 \\
\hline
\end{tabular}
\end{table}

The best results for linear and non-linear OCC models are obtained by appending our new regularization term $\gamma$ to SSVDD (see~\autoref{table: occperformance}). The performance of all three 
linear OCC models, 
namely: SSVDD$\gamma L_{kNN}$, SSVDD$\gamma L_{b}$ and SSVDD$\gamma L_{w}$, 
is nearly identical, bearing limited 
impact on performance metrics. 
On the other hand, among the non-linear OCC models, SSVDD$\gamma L_{kNN}$ demonstrates superior performance, outperforming the other two counterparts, namely SSVDD$\gamma L_{w}$ and SSVDD$\gamma L_{b}$, where the latter ranks the lowest in performance. 

\begin{figure*}[t]
\centering	
 \includegraphics[width=10cm, height=7.5cm]{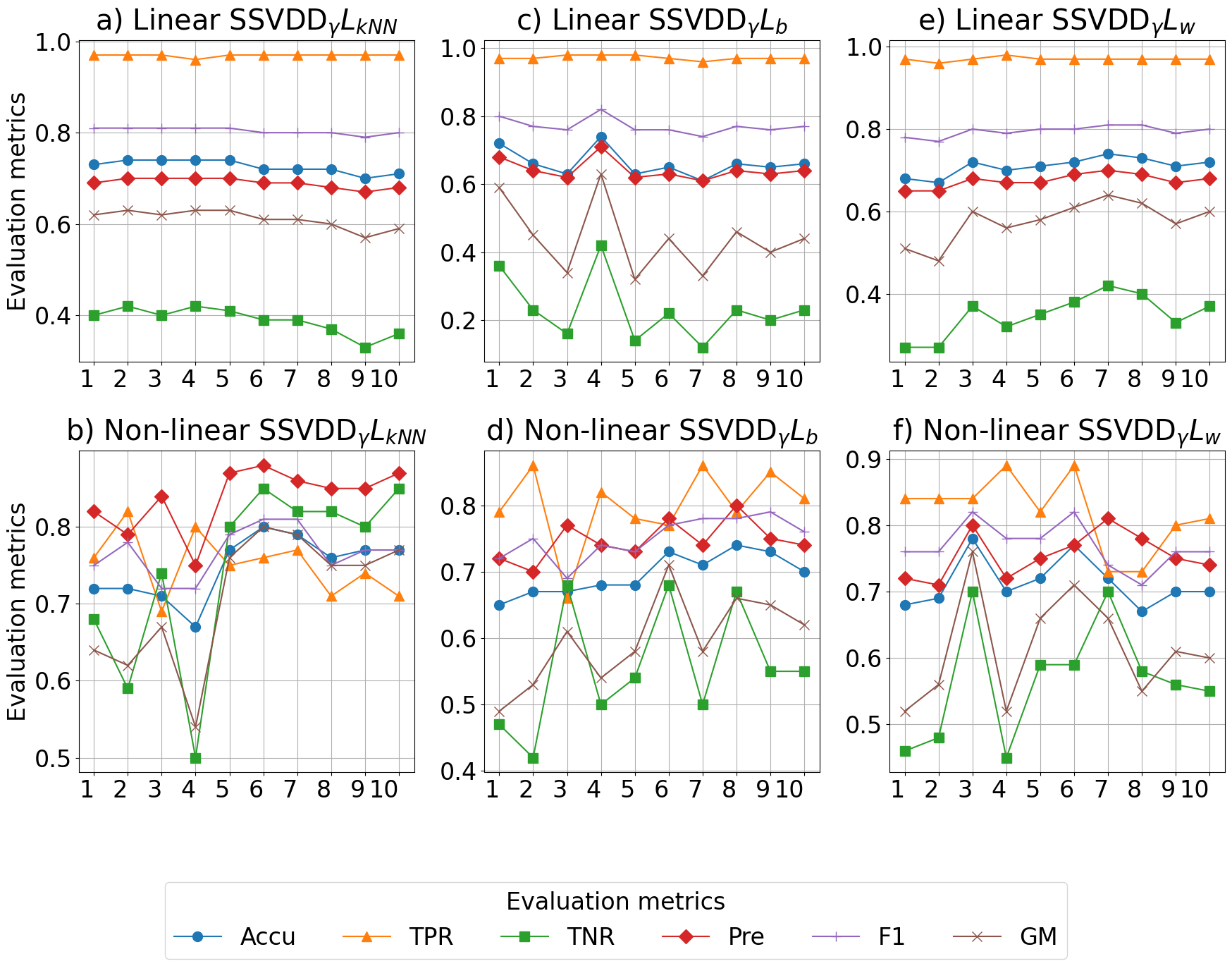}
	\caption{Evaluating SSVDD performance with the proposed regularization term $\gamma$, varying k for kNN, and different cluster values ($\mathcal{C}$) for $L_b$ and $L_w$.
 }
	\label{fig:occs}
\end{figure*}

Additional information about the use of $\gamma$ for SSVDD and its impact on performance metric can be found in~\autoref{fig:occs}~(a) for linear classification using kNN, and in~\autoref{fig:occs}~(b) for non-linear classifiers. Looking at the linear OCC models in~\autoref{fig:occs}~(a), the GM value remains steady at~0.57 to~0.63, and Accu falls within the range of~0.70 to~0.74 across various values of k for kNN, showing a stable performance level. All the performance metrics show stability, except TNR fluctuates, which tend to be on the lower side. 
Unlike the stable results in the linear classifier, the non-linear classifier shows a distinct pattern (see~\autoref{fig:occs}~(b)). The non-linear OCC displays variable performance, with GM values from~0.54 to~0.80 and Accu ranging from~0.67 to~0.80.

We also present the performance metrics for linear and non-linear SSVDD with the proposed regularization term $\gamma$, focusing on the between-cluster scatter Laplacian ($L_{b}$). As can be seen in~\autoref{fig:occs}~(c), and~\autoref{fig:occs}~(d), the choice of hyperparameter $\mathcal{C}$ in $L_b$ significantly impacts the performance of both linear and non-linear OCC models. The linear OCC models show varied performance across different values of $\mathcal{C}$ for $L_b$. For example, Accu ranges from~0.61 to~0.74, with the highest performance achieved at $L_{b=4}$, while precision falls within the range of 0.61 to 0.71. TPR fluctuates between~0.96 and~0.98, showing a reliable identification of positive instances. GM and TNR show a similar pattern, both displaying lower values. In contrast, non-linear OCC models show distinct behavior, with TPR at the top, varying between~0.66 and~0.86, followed by F1-score and Accu, both of which show a stable performance. However, GM and TNR, position at the last, show high variation. 
Following, we will analyze the results for within-cluster Laplacian ($L_w$). As can be seen in~\autoref{fig:occs}~(e), linear classifiers show varying performance across different $\mathcal{C}$ values for $L_w$. Accu ranges from~0.64 to~0.74 (at $L_w=7$), and TPR fluctuates between~0.96 and~0.98, indicating a consistent ability to identify positive instances correctly. Precision ranges from~0.65 to~0.70, and GM shows variations but generally remains between~0.56 and~0.64. The non-linear classifiers in~\autoref{fig:occs}~(f), show different behavior, with Accu ranging from~0.68 to~0.78. TPR varies between~0.73 and~0.89 and precision fluctuates between~0.71 and~0.81. It is noteworthy that at $L_w=3$, the non-linear OCC models achieve their highest F1-score and GM value.

\section{Conclusion}
\label{sec:conclusion}
Considering the significant impact of information sharing on social media, specifically on platform $\mathbf{X}$, our goal was to identify the trusted or untrusted $\mathbf{X}$ users. Our study provided insights into the effectiveness of OCC models in classifying political users on platform $\mathbf{X}$, through exploring OCC models. In addition, it included a novel regularization term for SSVDD.

In response to the research questions \textbf{\textit{RQ 1-3}}, our findings demonstrate the effectiveness of OCC models in identifying political $\mathbf{X}$ users as trusted or untrusted. The results consistently demonstrate that non-linear OCC classifiers outperform their linear counterparts. This paper provided brief insights on the recent improvements in OCC, notably the new paradigm of subspace learning for SVDD 
used to tackle the curse of dimensionality. Our study confirmed the potential 
of OCC performance optimization for political user identification through subspace learning. The proposed subspace-learning-based approach, particularly with the introduced regularization term for SSVDD, showcased superior performance compared to baseline models.

In the future, we will explore alternative kernel types and graph structures to enhance the performance further. 
Additionally, we aim to adapt the proposed regularization term to the Multi-modal Subspace Support Vector Data Description \cite{sohrab2021multimodal} framework and analyze its effectiveness over other application domains.

\bibliographystyle{unsrt} 
\balance
\bibliography{main} 

\end{document}